\documentstyle[preprint,aps,tabularx]{revtex}
\begin{document}


\title{Black Holes, Mergers, and the Entropy Budget of the Universe}

\author{Thomas W. Kephart}%
\address{
Department of Physics and Astronomy\\
Vanderbilt University,
Nashville, TN 37235, USA
}
\author{Y. Jack Ng}
\address{
Institute of Field Physics, Department of Physics and Astronomy\\
University of North Carolina,
Chapel Hill, NC 27599, USA
}

\maketitle

\begin{abstract}
 Vast amounts of entropy are produced in black hole formation, and the
 amount of entropy stored in supermassive black holes at the centers
 of galaxies is now much greater than the entropy free in the rest of
 the universe. 
 Either mergers involved in
 forming supermassive black holes are rare,
 or the holes must 
 be very efficient at capturing nearly all
 the entropy generated in the process.
 We argue that this information can be used to constrain supermassive black hole
 production, and may eventually provide a check on numerical results for
 mergers involving black holes.

\end{abstract}



In an ideal gas of $N$ identical particles, the entropy $S$ is an extensive
quantity proportional to $N$. Combining two ideal gases in equilibrium at
equal temperature and pressure with $N_{1}$ and $N_{2}$ particles
respectively, we find
\begin{equation}
S_{tot}\propto N_{1}+N_{2},
\end{equation}
and since $\Delta S=S_{tot}-S_{1}-S_{2}=0$, the process is reversible. In
the case of black hole (BH) thermodynamics, a black hole of
Schwarzschild radius $%
R=2M$ has entropy $S=\frac{1}{4}A$, where the area of the horizon is $A=4\pi
R^{2}=16\pi M^{2}$. Now combining two black holes, we find (assuming no
energy is lost in the process) $R=2(M_{1}+M_{2})$, and $S_{tot}=4\pi
(M_{1}+M_{2})^{2},$ so now $\Delta S=S_{tot}-S_{1}-S_{2}=8\pi M_{1}M_{2}$
and this process is therefore irreversible. $I.e$., entropy is produced.

On the other hand, if the
entropy of the new black hole is to be the same as the sum of the two
initial black holes, energy must necessarily be shaken off as they coalesce.
The lower bound for this is given by $S_{tot}=S_{1}+S_{2}$, or $%
M_{tot}^{2}=M_{1}^{2}+M_{2}{}^{2},$ and for $M=M_{1}=M_{2}$ we have
$M_{tot}=%
\sqrt{2}M$ and $\Delta E=2M-\sqrt{2}M\cong 0.59M.$

Considerable effort has been expended in studying the interactions of
black
holes with black holes, neutron stars, and other objects. While
results have been obtained for the ideal case of a head-on collision
between
two Schwarzschild black holes, the more typical astrophysical mergers of a
BH-BH binary, BH-neutron star pair or the capture of a
stellar mass object in a galactic core, will take longer to unravel. By a
typical merger, we mean the coalescence of two objects with different
masses, angular momenta, and magnetic fields spiralling together,
perhaps in
the environment of accretion disks or their distorted remnants. A fuller
understanding of such complicated but realistic situations will require
extensive numerical analysis.
In this Letter we consider the release of entropy into the inter-galactic
medium in the production of large black holes by mergers involving black
holes.  The result can be a useful
check on such numerical analyses.  It can also have bearings on structure
formation processes for galaxies, quasars, as well as for clusters\cite{Vala}.

The mass-energy that is thrown off in merger processes will contain some
amount
of
entropy, and the entropy to mass ratio $\rho _{s}^{out}$ of this
material
will depend on the parameters of the merging system, including time
($i.e$., on the progress of the merger). Although the entropy of the final
stable
black hole will have an initial entropy to mass ratio
$\rho _{s}^{in}$ linearly proportional to the
black
hole size as determined by black hole thermodynamics (we expect, for a
Schwarzschild black hole, $\rho
_{s}^{in}=\frac{\frac{1}{4}Area}{M}=2\pi R$), no such
result holds for $\rho _{s}^{out}(t)$. However, the
entropy
to mass ratio of the total system $\rho _{s}^{tot}=\rho _{s}^{in}+\rho
_{s}^{out}$ must be increasing in order to reach the Bekenstein value
(and
also because these processes are irreversible). Therefore we expect $\rho
_{s}^{out}$ to be increasing with time if the region from which it is to
be expelled
contains an incoherent mixture of this material with the matter that
will eventually fall
into the black hole. While this perspective is clearly naive, we take it
as
our working hypothesis, and postpone its refinement to later work.

Let us first
summarize the current state of knowledge concerning mergers.
The mergers of black hole-neutron star, neutron star-neutron star, and
black hole-black hole have all been studied in recent years, using
various methods (numerical simulations, analytic estimates, and
hydrodynamic simulations for neutron stars), in the
Newtonian or post-Newtonian approximations, or treated fully
relativistically
when feasible.  Compact binaries,
containing black holes or neutron stars, lose
energy in emitting gravitational radiation;
slowly they spiral towards each other until they reach the innermost stable
circular orbit when they start to coalesce and merge.\cite{Baum}
In the merger of
stellar mass black hole-neutron star binaries, a gas mass of a few $10^{-1}
M_{\odot}$ is left in an accretion torus around the black hole and
neutrinos are radiated.\cite{Janka}
Subsequent annihilations of neutrinos and antineutrinos
purportedly lead to gamma-ray bursts.
But while neutrino emission from the remnant of an inspiraling binary
neutron star following coalescence may be important for the cooling of the
remnant, it is negligible for the emission of angular momentum of
the merged objects; the consequent evolution of the remnant is dominated
by the emission of gravitational waves.\cite{BaSh}
Estimates for the Galactic merger rates range from $~ 10^{-5} yr^{-1}$
for neutron star mergers to an order of magnitude lower for black hole
and neutron star mergers.\cite{Yung}
On the stability of coalescing binary stars,
one group\cite{WMM} claims that massive neutron
stars, stable in isolation, individually collapse to black holes prior
to merger, while another group\cite{Shib} claims that the
tidal field from a binary
companion stabilizes a star against gravitational collapse.
Cosmological gamma-ray bursts are often
thought to be associated with gravitational
collapses of massive stars, but it has been suggested that the
binary neutron star merger scenario is actually more favored than
single stellar collapses.\cite{Tot}  Head-on
collisions of two equal mass, nonrotationing black holes for various
initial configurations have been studied using several independent
methods, with excellent agreement for total gravitational radiation
between numerical results and the analytic estimates.\cite{Ann1,Ann2}
The Binary
Black Hole Grand Challenge Alliance has used a three dimensional
numerical relativity code to study coalescing
black hole binaries.\cite{Abra}
We hope this Letter will be useful in providing a check on some of
the future works related to these results.

Since we are dealing with general relativity, we cannot think of $M$ as
proportional to
the number of particles in the system. In fact, since all the particles
in a black hole are
hidden inside the horizon, this type of counting loses meaning, as, for
instance, the number of baryons is the global quantum number, and gravity
violates all global symmetries. For a discussion of the implications 
see \cite{Holman:1992us}.

It is interesting to compare black hole entropy with the entropy of the
universe. Currently the entropy density is $s_{0}=2970\left(
\frac{T}{2.75K}%
\right) ^{3}\frac{1}{cm^{3}}$, so the entropy within our horizon today is
approximately
\begin{equation}
S_{0}=s_{0}(cH_{0}^{-1})^3 \cong 2.35\times 10^{87}\left(
\frac{Th^{-1}}{2.75K}%
\right) ^{3}
\end{equation}
On the other hand, the entropy of a Schwarzschild black hole is
\begin{equation}
S_{BH}=4\pi \left( \frac{M_{\odot }}{m_{pl}}\right) ^{2}\left( \frac{M}{%
M_{\odot }}\right) ^{2}\cong 1.05\times 10^{77}\left( \frac{M}{M_{\odot }}%
\right) ^{2}.
\end{equation}
Hence, a couple of hundred thousand solar mass black holes can contain
as much entropy as is free in the entire universe.
There is increasing evidence that supermassive black holes ($SMBH$s) exist
at the center of many galaxies, and that they are the sources which power
active galactic nuclei and quasars.\cite{Rees,Macc}
By now we know that a large fraction (at
least about 30\%) of galaxies contain such $SMBH$s with
masses $10^{6}M_{\odot }\lesssim M_{BH}\lesssim 10^{9}M_{\odot }$.
(Observations by the Hubble space telescope suggest that, e.g.,
our own galaxy has a $SMBH$ with mass $\sim 10^{6}M_{\odot}$ and M87
has one with mass $\sim 10^{9}M_{\odot}$.)
There are roughly $\sim 10^{11}$ galaxies in the universe.
Thus the black hole entropy
dominates all other sources of entropy.  $A$ $priori$ this is not a
problem, since as mentioned above black hole entropies are hidden behind
horizons. However, the formation of these supermassive black holes does
raise the question,
`How is it possible for them to form without considerable loss of
entropy to the
environment as one would expect from stellar mass black holes merging at the
cores of galaxies to form the supermassive black holes seen today?'

Let us give a very rough estimate of the total entropy stored in
$SMBH$s, where we assume they are
nonrotating. This gives,
\begin{eqnarray}
S_{BH}^{tot} &=&\sum_{gal.=1}^{N}4\pi \frac{M_{BH}^{2}(gal.)}{m_{pl}^2} \\
&\sim &3.2\times 10^{101}\times \left( \frac{M}{10^{7}M_{\odot }}\right)
^{2}\left( \frac{N}{10^{11}}\right)
\end{eqnarray}
where N is the number of galaxies within our horizion, and
$M_{BH}^{2}(gal.)$ is
the distribution of masses of the $SMBH$s at the galactic cores. (For some
galaxies this may be zero.)  To arrive at  the second line we have set all
values of $M_{BH}^{2}(gal.)$ to $M$. Comparing with $S_{0}$ and choosing $%
M=10^{7}M_{\odot }$ and $N=3 \times 10^{10}$ 
implies (very conservatively) that less than one
part in $%
\sim $10$^{14}$ of the black hole entropy could have escaped during
formation.  Because there is actually a distribution
of super heavy black hole
masses, a more accurate calculation of their entropy would take that into
account (and would be dominated by the heaviest $BH$s), though our order of
magnitude estimate will be more than sufficient for the arguments given
here.

Since big bang nucleosynthesis is highly dependent on the baryon-to-entropy
ratio, and since the present ratio is in agreement with big bang
nucleosynthesis ($BBN)$ calculations at the $\sim 0.1MeV$ scale, it appears
that any significant entropy production between the time of $BBN$ and today
can be ruled out. This means that in the formation of $SMBH$s, which
presumably did not start until the structure formation era at $z\sim 100$,
very little entropy could have escaped from the $BH$ forming regions. We
require that in order for the change in visible entropy $\Delta
S\lesssim 10S_0
$ from $BBN$ until today, the entropy released from $BH$ formation
satisfy $%
\Delta S_{BH}^{tot}\lesssim 10^{88}$ and so $\Delta S\lesssim 10^{-13} %
S_{BH}^{tot}$. Hence, the $BH$ formation must be extremely efficient at
keeping entropy hidden behind the horizon. We believe the easiest way to do
this is through nearly spherical collapse of large amounts of matter to
directly form $SMBH$s, and not through mergers that are potentially much
less efficient.
There is indirect support for this belief:  in the gravitational
collapse of a rotating supermassive star, it was found
recently\cite{Saijo}
that a supermassive black hole is formed coherently, with almost
all of the matter falling into the hole, leaving very little ejected
matter to form a disk, and strongly emitting gravitational waves.
This is also in line with arguments given by Silk and
Rees\cite{SilkRees} that $SMBH$s form in
protogalactic cores as quasarlike objects
before the epoch of peak galaxy formation, and argue against mergers as a
primary component of $SMBH$ formation. The quasars then had profound effects
on star formation. We conclude it is most likely that $SMBH$s formed
in gravitational collapse events in protogalactic cores at or before galaxy
formation.  Alternatively, if $SMBH$s were to have formed via mergers, as seems
less likely, then the mergers must have been extremely efficient at
hiding all the entropy produced, and thus would certainly be more
probable to have taken place in
regions of low ambient density.

We can use the numerical results of
Anninos $et\ al$. \cite{Ann1,Ann2} to get
an estimate of energy radiated and entropy produced in $BH$ mergers. For
two equal-mass $M$, uncharged and non-rotating $BH$s that merge via a
head-on collision, the total energy radiated in gravitons is $E\cong
0.002M$%
. This is in agreement with earlier analytic results which gave
\begin{equation}
E\cong 0.0104\frac{\mu ^{2}}{M}  \label{headon}
\end{equation}
where $\mu =\frac{mM}{m+M}\ $is the reduced mass for two $BH$s of mass $m$
and $M$, and so for $m=M$, gives $E\cong 0.0026M$. More mass is
expected to
be radiated in collisions with non-zero impact parameter and for rotating
and for charged $BH$s, or if the $BH$s are surrounded by a medium or
accretion disks. Hence we take this as an approximate lower bound.

For wavelength $\lambda \sim R=2M$ the number of gravitons produced is
$%
N\sim \frac{E}{\varepsilon _{\gamma }}\sim 5\times
10^{72}\left(\frac{M}{M_{\odot}}\right)^2$. But,
although an enormous number of gravitons are produced
in $BH$ mergers, they would have a high degree of coherence, and so only a
small
amount of entropy could be released in the process. Where we expect a
large
entropy production is in mergers of $BH$s with neutron stars, and
mergers
that take place within dense media, $e.g$. involving dense protogalactic
cores or overlapping and/or distorted accretion disks. Then much of the
energy released could be in photons, with very little coherence due to
local
dissipation. The two merging $BH$s or neutron star and $BH$ will
coalesce
into a single but very perturbed black hole. The evolution into one
black
hole with a single horizon is still not well understood, and is the
subject
of ongoing numerical work. Hence the potential for entropy production
from
mergers is enormous and can, given sufficient numerical work, eventually
be
used to bound the number and types
of mergers allowed since the time of the beginning of the growth of
density
perturbations $z\lesssim 100$. Indeed, as quasar formation peaked at
around $%
z\sim 3$, and star
formation peaked around $z\sim 1.5$, this also suggests fewer mergers and
larger
intitial $BH$s ($QSO$s). (In fact, $SMBH-SMBH$ mergers may also be
constrained by the
fact that life exists on this planet, as accretion of dwarf galaxies and
subsequent mergers would act as super gamma
ray
bursts. The resulting luminosity would be greater than $L_{\odot}$ within
$\sim 3Mpc$, and lasts
for
minutes to hours, so within the galaxy this could be fatal. However, while this
thought is intriguing, it may be academic since it
is known that even if
galaxies merge, their $SMBH$s have a low probability of merging too, and
orbits of several $pc$ are
stable for times longer than the age of the Universe.)

Conclusions: Under some modest assumptions we have concluded that either
mergers
involving black holes that generate larger black holes are rare, or they
must capture nearly all the entropy generated in the process. This has
implications for galaxy formation and cosmological models. It can also
serve
as a guide and check on further numerical work on mergers. While our
results
are necessarily imprecise, the fact that less than one part in $\sim
10^{13}$
of the entropy stored in black holes has been released into the
intergalactic environment is provocative.

%
%
%
We thank Chuck Evans for useful discussions and for pointing us to a
number of references.
This work was supported in part by the US Department of Energy
under Grants No. DE-FG02-97ER-41036 and DE-FG05-85ER-40226, and by
the Bahnson Fund of the University of North Carolina.

\end{document}